# A novel design of all-optical half adder using a linear defect in a square lattice rod-based photonic crystal microstructure

Saleh Naghizade[1], and Hamed Saghaei[2,*]

[1] Young Researchers and Elite Club, Tabriz Branch, Islamic Azad University, Tabriz, Iran
[2] Department of Electrical Engineering, Shahrekord Branch, Islamic Azad University, Shahrekord, Iran

*Corresponding Author: h.saghaei@iaushk.ac.ir

**Abstract**

In this paper, we propose an all-optical high-speed half adder based on linear defects in a photonic crystal (PhC) structure composed of silicon rods. The proper design of half adder results in no need to increase the intensity of the input optical signal for the appearance of the nonlinear Kerr effect, which leads to diverting the incoming light toward the desired output. The proposed device consists of four optical waveguides and a defect in a square lattice PhC. Two famous plane wave expansion and finite difference time domain methods are used to study and analyze photonic band structure and light propagation inside the PhC, respectively. The presented structure, the ON-OFF contrast ratios for Sum and Carry, are 16dB and 14dB, respectively. Our simulation results reveal the proposed half adder has a maximum delay time of 0.7 ps with a total footprint of 158 μm$^2$. Due to very low delay time, high contrast ratio, and small footprint that are more crucial in modern optoelectronic technologies, this structure can be used in the next generation of all-optical high-speed central processing units.

## 1. Introduction

Photon based devices have been the focus of researchers in recent decades. It is due to their high processing speed, small area, and low power consumption. Organizing the optical communication systems based on high-speed devices is one of the human concerns that is growing at high speed. Information or data processing is very important in optical networks. For realizing an all-optical data processing system, we need all-optical logic-based devices. One important device for implementing optical data processing is an all-optical half adder because all four basic mathematical operations such as addition, subtraction, multiplication, and division can be performed using optical half adders [1–5]. Photonic crystals (PhCs) can play key roles in the all-optical systems and networks [6,7]. The existence of photonic band gaps (PBGs) in the certain frequency range of these periodic structures gives them the ability to confine and control the propagation of light at desired waveguides [8–10]. Therefore, many optical devices such as optical filters [11–13], PhC fibers [14–18], demultiplexers [19–24], switches [10,25,26] logic gates such as NOT, AND, OR, NAND, encoder, and decoder [27–29] and analog to digital converters (ADCs) [30–34] have been designed using this property of the PhCs. Recently, all-optical half adders have been designed and investigated based on PhCs. Most of the previously proposed logic gates can be classified into two main categories: nonlinear Kerr-effect-based structures and linear phase-difference-based structures. The Kerr-effect base structure requires high input power to appear the nonlinear Kerr effect. However, in this case, damaging the structure is probable. It is due to using a high intensity of the input optical signal. Researchers presented different structures of half adder so far. Jiang et al [5] proposed an optical half adder using self-collimated beams inside 2D PhCs. The presented structure was realized by introducing two line defects inside the structure, which can work as a power splitter. Ghadrdan et al [35] proposed a PhC-based half adder by combining AND and XOR gates, inside a 2D PC structure. Xavier et al. [36] presented another PhC-based half adder by combining line defects and self-collimated beams inside 2D PhC. The proposed structure was composed of AND and XOR logic gates. Rahmani and Mehdizadeh [37] proposed an optical half adder using three nonlinear ring resonators inside 2D PhC. The nonlinear resonators were created by adding some nonlinear rods to the PhC-based ring resonator structure. The maximum rise and fall times of their proposed structure are about 1.5 and 1 ps, respectively. Several optical half-adder devices have been reported with other authors so far [4,28,38,39]. In this paper, we design a linear phase-difference-based half adder structure with a low input optical power. It has a good capability to separate logics 0 and 1 at the outputs. To confine and guide the light in the waveguides and defect region, we need the wavelengths which are in the PBG of the PhC. Thus, the proposed structure is simulated at the c-band communication window (at 1550 nm). One property of this half adder is a relatively large difference between the two logical levels. It has also low delay time, low input power, and small footprint. All this results in an error reduction for a high-speed data processing system. The rest of the paper is organized as follows: In Section 2, we present the design procedure of the proposed half adder. Numerical results accompanied by discussions are given in Section 3, and finally, the paper is closed by the conclusion in Section 4.

## 2. PhC structure and proposed half adder

A half adder is used to add two single binary digits and provide the output plus with a carry value. We assign symbols *X* and *Y* to the two inputs and Sum and Carry to the outputs. The common representation uses an XOR logic gate and an AND logic gate. In fact, Sum=$X \oplus Y$ and Carry=$X \cdot Y$ where $\oplus$ sign stands for XOR and · sign stands for AND. The Carry output is 1 only when both inputs are 1. The Sum output represents the least significant bit of the sum. In fact, half adder is the basic building block of an arithmetic logic unit which is used in every optical central processing unit (CPU) to provide computational operators.

Figure 1 represents the block diagram and truth table of a half adder. As can be observed in the figure, a half adder has two inputs and two outputs. The output ports are S and C, where S stands for the sum, and C stands for the carry. Also, X and Y are the input ports of the half adder. To design an all-optical half adder, we employ a 21*21 array of dielectric rods with a square lattice in air background. The refractive index of dielectric rods is 3.46. The radius of rods is r=0.2*$a$, where *a* is the lattice constant of the PhC structure that is 600 nm. We calculated the band diagram of the fundamental structure using the plane wave expansion (PWE) method [40]. According to Fig. 2, there are two PBGs in TM mode (blue color areas). The first PBG in TM mode that is 0.285<a/λ<0.418, has the appropriate frequency range for our goals. By choosing the lattice constant of *a* = 600 nm, the PBG will be at 1435 nm<λ< 2105 nm, which completely covers the wavelength range of the C-band communication window.

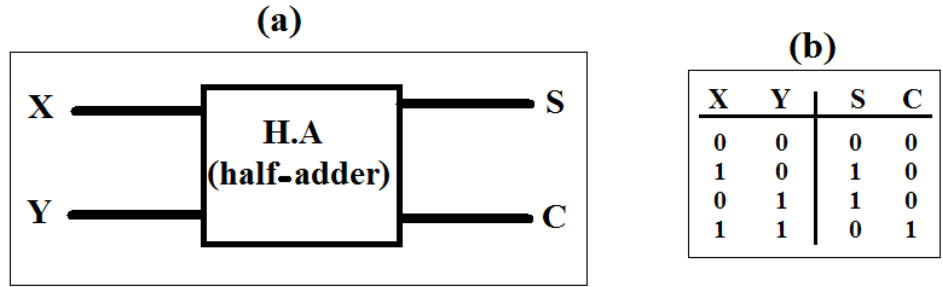

Fig. 1. (a) Block diagram, and (b) truth table of an all-optical half adder.

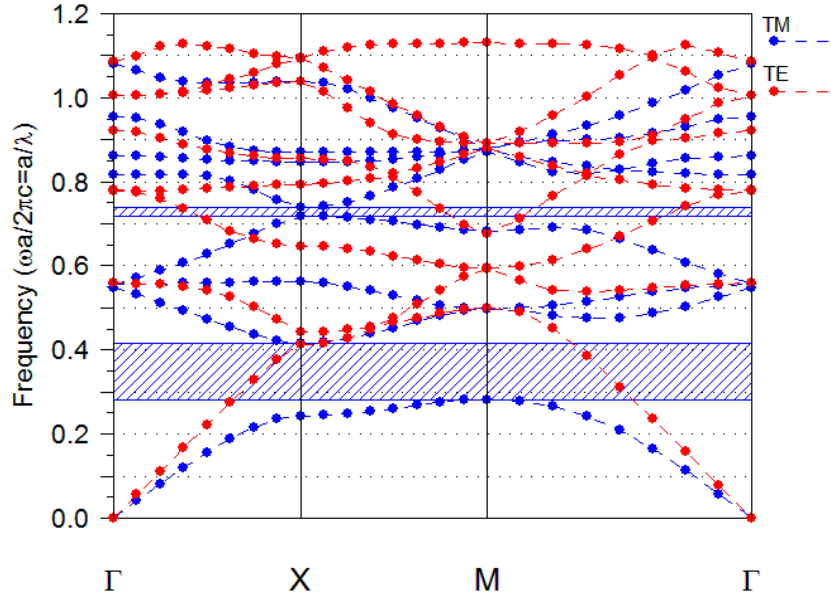

Fig. 2. Photonic TM- and TE-mode band diagrams of a fundamental square lattice PhC structure with rod radius of r=120 nm and lattice constant of $a$=600 nm.

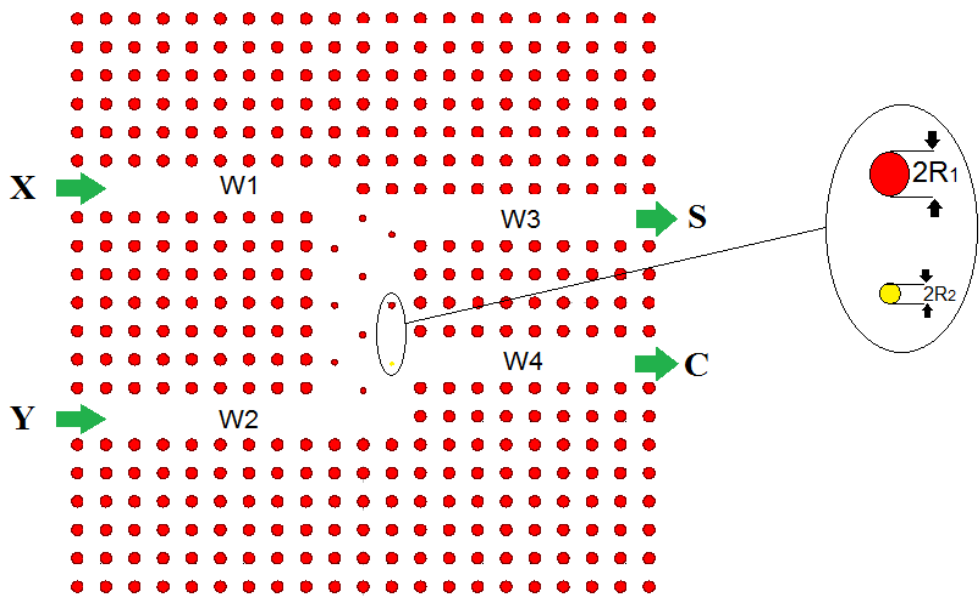

Fig .3. The proposed all-optical half adder in a square lattice PhC composed of silicon rods with r=120 nm, $R_1$=60 nm, and $R_2$=30 nm and $a$=600 nm.

For all-optical half-adder design in a 21×21 array of dielectric rods, four optical waveguides and one resonant cavity are created in certain regions of the PhC structure shown in Fig. 3. Indeed, the combination of W1, W2, W3, and W4 waveguides with resonant cavity build our optical half adder. The defect region contains 10 folds dielectric rods in which the radius of 9 folds of defect rods ($R_1$) is 60 nm. The radius of the last defect rod ($R_2$) shown by yellow is 30 nm. It is located at the input of W4. The end paths of W3 and W4 are the S and C ports of the proposed half adder, respectively. X and Y are the input ports of the half adder at the beginning of the W1 and W2, respectively.

## 3. Numerical results and discussions

In order to simulate the proposed structure, we used the finite-difference time-domain (FDTD) method [41,42]. Performing a 3D simulation for studying the proposed structure requires a powerful computer as well as a long simulation time. Due to time and memory limitations, we applied the effective refractive index method to reduce 3D simulations into 2D simulations with acceptable accuracy [43]. The proposed half adder has 2 input ports, so we have 4 different input states. Therefore, we employed optical waves with a central wavelength of 1550 nm at the input port. All cases of the half adder are shown in Fig. 4 and classified as follows:

**Case 1:** When both input ports (X and Y) are OFF, there is no optical power inside the structure; therefore, both output ports (S and C) will be OFF.

**Case 2** and **Case 3:** When one of the input ports (either X or Y) is ON due to wavelength matching between the resonant mode of the resonant cavity and input signal, the resonant cavity will couple the optical beams into W3. Therefore, in these cases, S will be ON and C will be OFF.

**Case 4:** When both input ports (both X and Y) are ON, the resonant cavity will couple optical beams into W4. Therefore, in this case, S will be OFF and C will be ON.

Comparing the obtained results with the truth table shown in Fig. 1(b), it is confirmed that the proposed structure can operate as an all-optical half adder. The normalized output of the proposed structure is shown in Fig. 5. As shown in Fig. 5(a), when the input port of X is ON, the normalized intensities of S and C ports are 75% and 5%, respectively. In this case, the delay time is about 0.7 ps. Figure 5(b) demonstrates, when the Y input port is ON, the normalized intensities of S and C ports are 85% and 2% respectively. In this case, the delay time is about 0.8ps. Figure 5(c) shows when both input ports are ON, the normalized intensities of S and C ports are 2% and 125%, respectively. In this case, the delay time is about 0.7 ps. The obtained results show that our proposed structure has a shorter delay, a lower input power (equal to 1 W/µm$^2$ because we don't use nonlinearity) and a smaller footprint in comparison to previously reported structures. Also, Considering these results, the ON-OFF contrast ratios ($10 \times Log(P_{ON}/P_{OFF})$) for both SUM and CARRY ports are 16 dB and 14 dB, respectively. Also, according to the presented diagrams, the maximum delay time is about 0.7 ps. We considered the time required for the output port to reach its steady-state as the delay time.

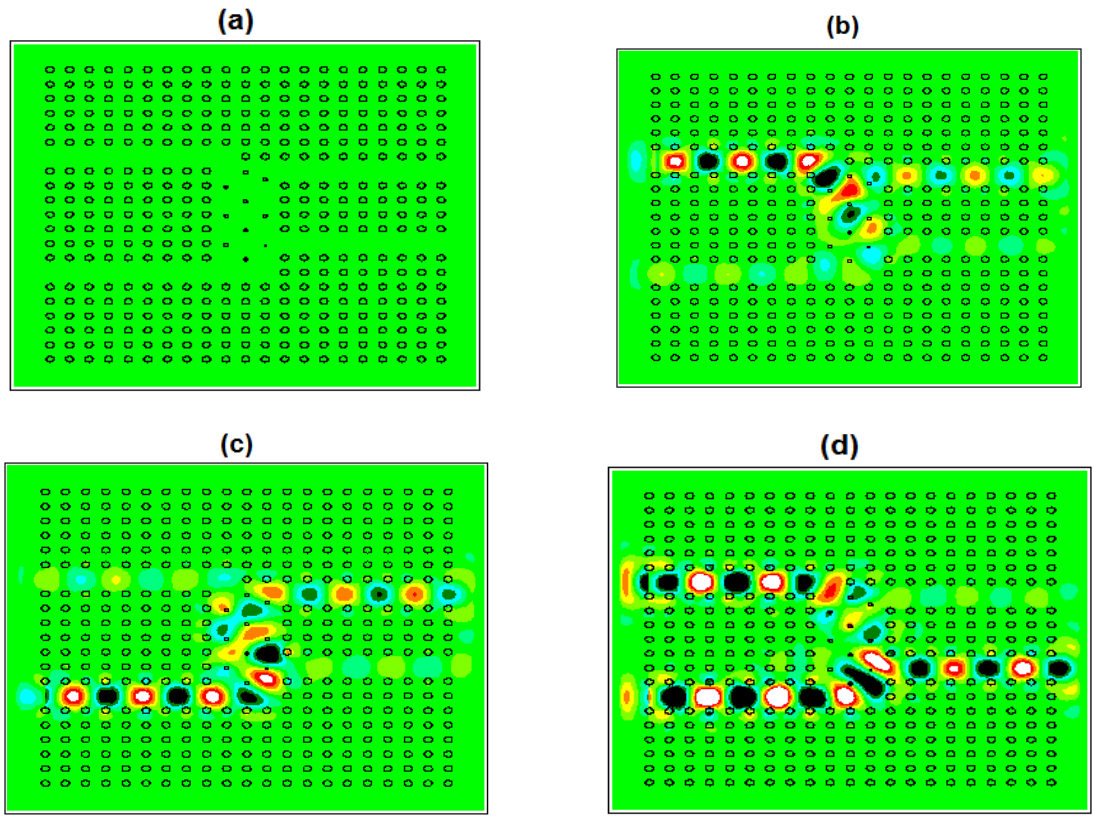

Fig. 4. Light propagation inside the proposed half adder for (a) Case 1 (b) Case 2, (c) Case 3, and (d) Case 4.

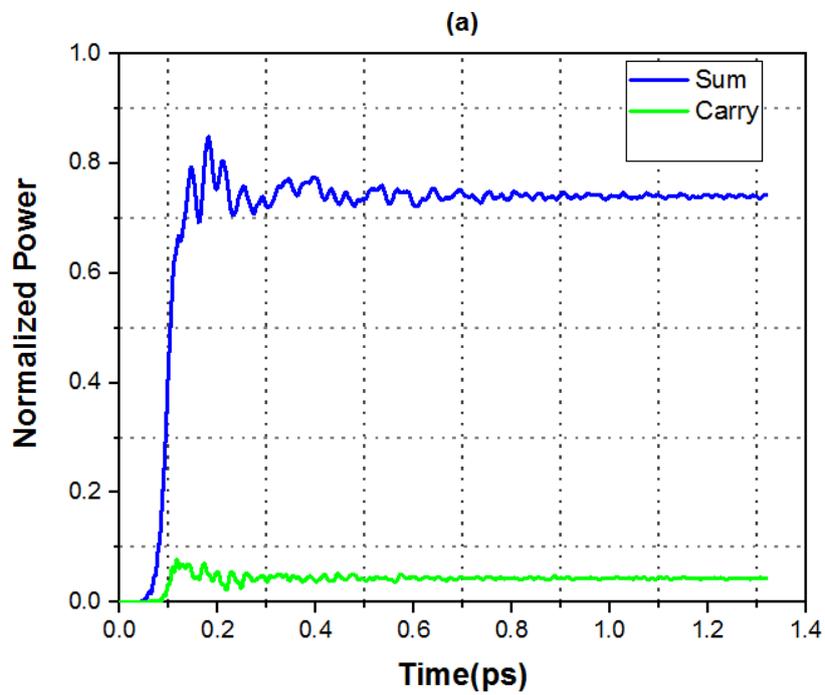

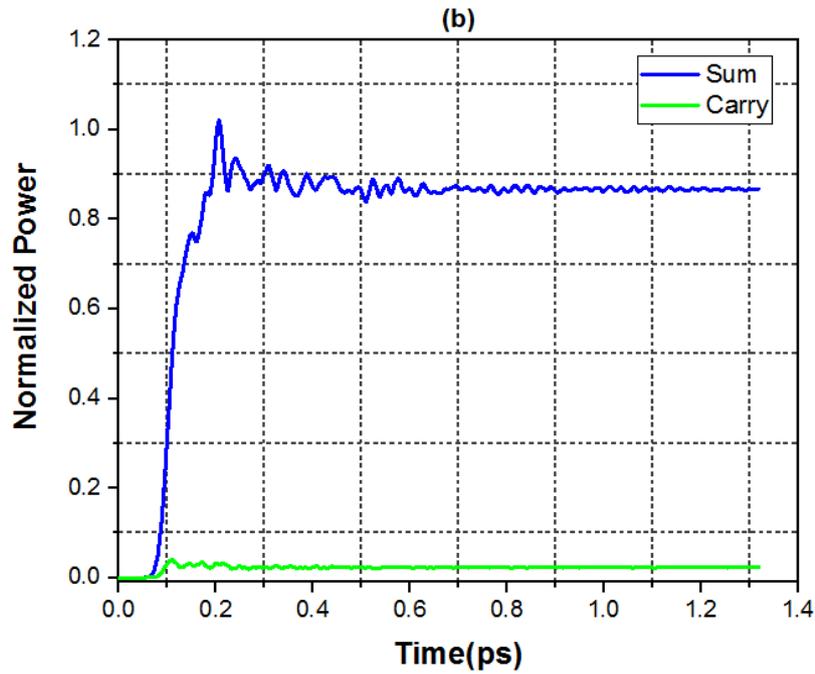

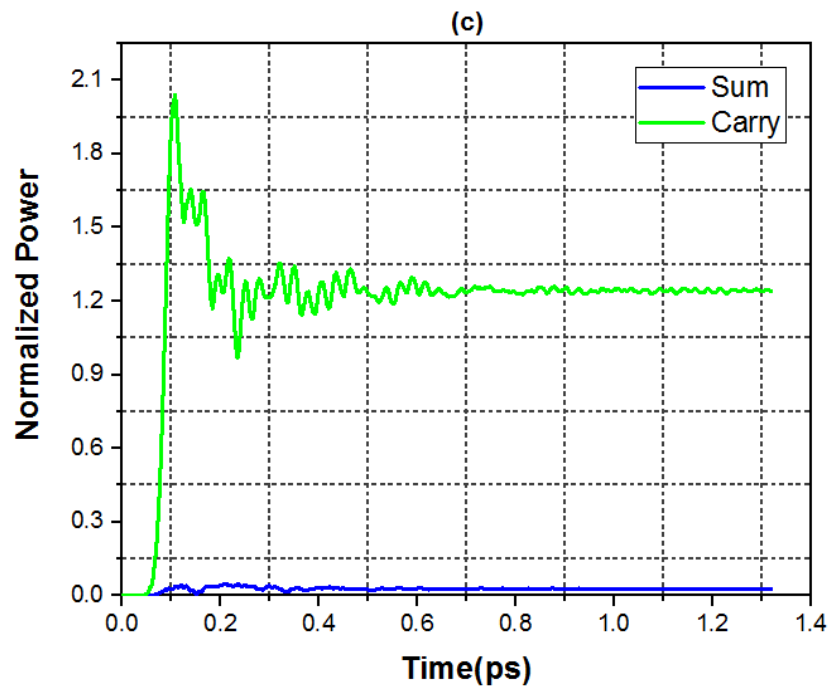

Fig. 5. Normalized output power versus time of the proposed half adder for (a) Case 2 (b) Case 3, and (c) Case 4.

Table 1 represents the comparison of the proposed device performance with other published papers [4,5,35–38,44,45]. As can be observed the minimum transmission of logic 1 and the maximum transmission of logic 0 are 4% and 75%, respectively. The calculations also demonstrate the proposed half adder has a delay of 0.7 ps due to having a small footprint of 158 μm$^2$.

## 4. Conclusion

In this paper, we designed an ultrafast all-optical half adder based on a photonic crystal microstructure in an area of 158 $\mu m^2$. The photonic band diagram was calculated using the plane wave expansion method for TE and TM polarization modes. We also studied the light propagation in the device via the finite-difference time-domain method and calculated the outputs for different values of input ports. One of the most important advantages of our structure compared to similar studies was the non-use of high nonlinear dielectric rods. This resulted in no need to increase the input power to divert the incoming light emission to the desired output. Simulations revealed the minimum transmission of logic 1 and the maximum transmission of logic 0 are 4% and 75%, respectively. The calculations also demonstrated the proposed half adder has a delay of 0.7 ps due to having a small area. All this makes it an appropriate building block of photonic integrated circuits, especially in the next generation of optical computers.

Table 1: Comparison of the proposed half adder with other published papers.

| Works | Method | Min power for logic 1 | Max power for logic 0 | time delay | footprint |
|---|---|---|---|---|---|
| Ref. [5] | self-collimation | 50% | 7% | - | - |
| Ref. [35] | nonlinear | 81% | 22% | 0.85 ps | 168 $\mu m^2$ |
| Ref. [36] | self-collimation | 73% | 24% | - | 169 $\mu m^2$ |
| Ref. [37] | nonlinear | 100% | 0% | 1 ps | - |
| Ref. [4] | nonlinear | 96% | 4% | 3.6 ps | 250 $\mu m^2$ |
| Ref. [28] | linear | 71% | 22% | - | - |
| Ref. [38] | nonlinear | 95% | - | 0.91 ps | - |
| Ref. [44] | linear | 95% | 19% | 4 ps | 1056 $\mu m^2$ |
| Ref. [45] | linear | 45% | 19% | 0.48 ps | 171 $\mu m^2$ |
| This work | linear | 75% | 5% | 0.7 ps | 158 $\mu m^2$ |